\documentstyle[preprint,floats,prd,aps]{revtex}
\begin{document}
%

\draft
\title{$B\to K\gamma\gamma$ via intermediate $\eta'$} \author{Mohammad R. Ahmady
\footnote{Email address: mahmady2@julian.uwo.ca}}
\address{
Department of Applied Mathematics, University of Western Ontario\\ London, Ontario, Canada N6A 5B7 \\}

\date{August 1999}
\maketitle
\begin{abstract}
We examine our previous conjecture that the $\eta'$ intermediate resonance has the dominant role in the long distance contributions to $B$ decay into two photons and a strange final state hadron.  We calculate the branching ratio of the exclusive $B\to K \eta'\to K\gamma\gamma$ decay using the nonspectator mechanism for $\eta'$ production in charmless hadronic $B$ decays. It is shown that the obtained branching ratio $B^{(\eta')}(B\to K\gamma\gamma )\approx 8.7\times 10^{-7}$ is more than twice as large as the $\eta_c$ contribution to this decay mode.  
\end{abstract}
%

\newpage
The need to understand and isolate the long distance (LD) background to rare $B$ decays is of utmost importance.  Some rare decay modes are actually dominated by LD contributions and therefore, to search for such short distance (SD) effects like "new physics", one needs to find observables and regions of various decay distributions where the LD interference is minimal.

In this paper, we examine the contribution of the $\eta'$ meson to the LD background in the exclusive $B\to K\gamma\gamma$ decay.  In fact, in a previous work on the intermediate pseudoscalar resonance contributions to inclusive $B\to X_s\gamma\gamma$ decay mode, we had conjectured that $\eta'$ might have a more significant role in generating LD background than $\eta_c$. \cite{aks98}.  This was due to the observation of the CLEO Collaboration of an unexpectedly large branching ratio for the inclusive $B\to X_s\eta'$ and its exclusive counterpart $B\to K\eta'$\cite{s,b,bw}.  A rough estimate of the ratios of the branching fractions shows that
\begin{equation}
\frac{B^{(\eta')}(B\to X_s\gamma\gamma )}{B^{(\eta_c)}(B\to X_s\gamma\gamma )} \approx\frac{B(B\to X_s\eta' )B(\eta'\to\gamma\gamma)}{B(B\to X_s\eta_c )B(\eta_c\to\gamma\gamma)}\approx 6\;\; ,
\end{equation}
and
\begin{equation}
\frac{B^{(\eta')}(B\to K\gamma\gamma )}{B^{(\eta_c)}(B\to K\gamma\gamma )} \approx\frac{B(B\to K\eta' )B(\eta'\to\gamma\gamma)}{B(B\to K\eta_c )B(\eta_c\to\gamma\gamma)}\approx 5\;\; ,
\end{equation}
where $B(B\to X_s\eta_c )\approx 8.7\times 10^{-3}$ and $B(B\to K\eta_c )\approx 1.1\times 10^{-3}$ has been used\cite{ak,am}.  In order to actually calculate the left-hand-side of the Eqs. (1) and (2), one needs to have a model for the fast $\eta'$ production in charmless nonleptonic $B$ decays which can produce the CLEO experimental data.  Various mechanisms for this process and related issues have been discussed in the literature\cite{as,aks'98,hz,yc,ht,kp,aek}.  Our proposed nonspectator gluon fusion model in Ref. \cite{aks'98}, which is closely related to the gluon fragmentation mechanism of Atwood and Soni \cite{as}, has the advantage of fomulating both inclusive and exclusive $B$ decays to $\eta'$ and the final state hadron containing a strange quark in the same context.  Here, we use this mechanism to calculate the LD contribution of the intermediate $\eta'$ resonance to the exclusive $B\to K\gamma\gamma$ decay mode and compare it to the background generated by $\eta_c$, i.e. $B^{(\eta_c)}(B\to K\gamma\gamma )$.

Using the dominant chromo-electric component of the OCQ penguin diagram, we obtain the following nonspectator effective Hamiltonian \cite{aks'98,ak99}  
\begin{equation}
H_{eff}=iCH(\bar s\gamma_\mu (1-\gamma_5)T^ab)(\bar q\gamma_\sigma T^a q)\frac{1}{p^2}\epsilon^{\mu\sigma\alpha\beta}q_\alpha p_\beta\;\; ,
\end{equation}
where
\begin{equation}
C=\frac{G_F}{\sqrt{2}}\frac{\alpha_s}{2\pi}V_{tb}V_{ts}^*[E(x_t)-E(x_c)]\;\; ,
\end{equation}
and $p$ and $q$ are the momenta of the two gluons which combine to produce the $\eta'$ meson.
The coefficient function $E$ is defined as
\begin{equation}
\nonumber
E(x_i)=-\frac{2}{3}Lnx_i+\frac{x_i^2(15-16x_i+4x_i^2)}{6{(1-x_i)}^4}Lnx_i+
\frac{x_i(18-11x_i-x_i^2)}{12{(1-x_i)}^3}\;\; ,
\end{equation}
where $x_i=m_i^2/m_W^2$ with $m_i$ being the internal quark mass\cite{il}.   $H$ is the form factor parametrizing the $g-g-\eta'$ vertex  
\begin{equation}
A^{\mu\sigma}(gg\to\eta')=iH(q^2,p^2,m_{\eta'}^2)\delta^{ab}
\epsilon^{\mu\sigma\alpha\beta}q_\alpha p_\beta\;\; .
\end{equation}
Using the decay mode $\psi\to\eta'\gamma$, $H(0,0,m_{\eta'}^2)$ is estimated to be approximately 1.8 GeV$^{-1}$\cite{as}.  A re-arrangement of Eq. (1) via Fierz transformation and using the definition of the decay constants for the $B$ and $K$ mesons in the context of factorization, we can express the matrix element for $B\to K\eta'$ decay as the following:
\begin{equation}
<\eta' K|H_{eff}|B>=-i\frac{2CHf_Bf_K}{9p^2}\left (p_B.qp_K.p-p_B.pp_K.q\right )\;\; .
\end{equation}
Inserting $p^2\approx -\Lambda_{\rm QCD}^2\approx -0.3^2$ GeV$^2$, $p_0=0.3$ GeV ($p_0$ is the energy transfer by the gluon emitted from the light quark in the $B$ meson rest frame), $f_B=0.2$ GeV and $\alpha_s=0.2$ in the decay rate derived from Eq. (7) results in the following exclusive branching ratio:
\begin{equation}
B(B\to K\eta' )=7.1\times 10^{-5}\;\; ,
\end{equation}
which is in good agreement with the experimental data\cite{s,b}.
The matrix element in Eq. (7) along with the transition amplitude for $\eta'\to\gamma\gamma$, which can be written in the form
\begin{equation}
A(\eta'\to\gamma\gamma )=N\epsilon^{\mu\nu\alpha\beta}\epsilon_\mu (p_1)\epsilon_\nu (p_2)p_{1\alpha}p_{2\beta}\;\; ,
\end{equation}
where $N=i8\sqrt{\pi\Gamma (\eta'\to\gamma\gamma )}/m_{\eta'}^{3/2}$ and, $\epsilon (p_1)$ and $\epsilon (p_2)$ are the polarizations of the two final state photons with momenta $p_1$ and $p_2$, respectively, lead to the amplitude for the intermediate $\eta'$ resonance contribution to the exclusive $B\to K\gamma\gamma$ decay
\begin{eqnarray}
\nonumber
A^{(\eta')}(B\to K\gamma\gamma )&=&-i\frac{2CHf_Bf_K}{9p^2}\left (p_B.qp_K.p-p_B.pp_K.q\right )\\
&\times &\frac{N\epsilon^{\mu\nu\alpha\beta}\epsilon_\mu (p_1)\epsilon_\nu (p_2)p_{1\alpha}p_{2\beta}}{{(p_1+p_2)}^2-m_{\eta'}^2+im_{\eta'}\Gamma_{\eta'}}\;\; .
\end{eqnarray}
$\Gamma_{\eta'}\approx 0.2$ MeV is the total width of the $\eta'$ meson \cite{pdg}.  As a result, the differential decay rate can be obtained in a straightforward manner:
\begin{eqnarray}
\nonumber
\frac{d\Gamma^{(\eta' )}(B\to K\gamma\gamma )}{ds}&=&\displaystyle\int_0^\pi\frac{\overline{{\vert A\vert}^2}}{512\pi^3m_B^3}\sqrt{({(m_B-m_K)}^2-s)({(m_B+m_K)}^2-s)}\sin{\theta} d\theta \\
\nonumber &=&\frac{C^2H^2f_B^2f_K^2N^2}{12^5\pi^3m_B^3{(p^2)}^2}\left [ 3p_0^2{(s(m_B+E_K)-(m_B^2-m_K^2)(m_B-E_K))}^2 \right . \\
\nonumber &+&\left . {(m_B^2-m_K^2+s)}^2{\vert\vec p_K\vert}^2(p_0^2-p^2)\right ]\frac{s^2}{{(s-m_{\eta'}^2)}^2+m_{\eta'}^2\Gamma_{\eta'}^2}\\
&\times & \sqrt{({(m_B-m_K)}^2-s)({(m_B+m_K)}^2-s)}\;\; ,
\end{eqnarray}
where the parameter $\theta$ in the first line is the angle of the K meson and $E_K$ and $\vec p_K$ are its energy and momentum in the rest frame of the decaying B particle.  $s={(p_1+p_2)}^2$ is the invariant mass of the two photons.  Using the same parameter values which resulted in the decay rate in Eq. (8), we observe that the peak at $s=m_{\eta'}^2$ leads to a large LD contribution to the exclusive decay mode
\begin{equation}
B^{(\eta')}(B\to K\gamma\gamma )\approx 8.7\times 10^{-7}\;\; .
\end{equation}
To compare the above result with the LD effect of the intermediate $\eta_c$ resonance, we calculate $B^{(\eta_c)}(B\to K\gamma\gamma )$ derived from the following amplitude
\begin{equation}
A^{(\eta_c)}(B\to K\gamma\gamma )=\frac{C'N'f_{\eta_c}F_0({(p_1+p_2)}^2)(m_B^2-m_K^2)}{{(p_1+p_2)}^2-m_{\eta_c}^2+im_{\eta_c}
\Gamma_{\eta_c}}\epsilon^{\mu\nu\alpha\beta}\epsilon_\mu (p_1)\epsilon_\nu (p_2)p_{1\alpha}p_{2\beta}\;\; ,
\end{equation}
where
\begin{eqnarray}
\nonumber
C'&=& \frac{G_F}{\sqrt{2}}V_{cb}^*V_{cs}\left (C_2(\mu)+\frac{1}{3}C_1(\mu)\right )\;\; ,\\
\nonumber
N'&=& i\frac{8\sqrt{\pi\Gamma (\eta_c\to \gamma\gamma )}}{m_{\eta_c}^{3/2}}\;\; ,
\end{eqnarray}
and with the Wilson coefficients $C_1$ and $C_2$ evaluated at the scale $\mu\approx m_b$.  $F_0$ is the form factor parametrizing $B\to K$ hadronic matrix element and $f_{\eta_c}$ is the decay constant of $\eta_c$ pseudoscalar meson.  Consequently, the differential decay rate for $\eta_c$ mediated $B\to K\gamma\gamma$ can be written as
\begin{equation}
\frac{d\Gamma^{(\eta' )}(B\to K\gamma\gamma )}{ds}=\frac{{C'}^2{N'}^2{F_0(s)}^2}{1024m_B^3}\frac{s^2\sqrt{{(m_B^2-m_K^2-s)}^2-4sm_K^2}}{{(s-m_{\eta_c}^2)}^2+m_{\eta_c}^2\Gamma_{\eta_c}^2}\;\; .
\end{equation}
We have neglected the suppression form factor for $\eta_c\to\gamma\gamma$ transition when $\eta_c$ is far off its mass-shell mainly because the total decay rate is not affected significantly \cite{aks98}.  Using a monopole form for the form factor with $F_0(0)=0.38$ obtained in the BSW model \cite{bsw}, and $C_1+1/3C_2=0.155$, which is adopted from a next-to-leading order calculation \cite{buras}, along with $f_{\eta_c}\approx 0.48$ GeV result in our estimate of the branching fraction via $\eta_c$:
\begin{equation}
B^{(\eta_c)}(B\to K\gamma\gamma )\approx 3.3\times10^{-7}\;\; .
\end{equation}
A comparison between Eqs. (12) and (15) reveals that the LD contribution via $\eta'$ is more than twice as large as the one mediated by $\eta_c$.  The total background due to these resonances is basically the sum of the two contributions, i.e. $B^{LD}(B\to K\gamma\gamma )\approx 1.2\times 10^{-6}$, as there is no appreciable interference between the two amplitudes of Eqs. (10) and (13).  Considering the estimated SD branching ratio $B^{SD}(B\to X_s\gamma\gamma )\approx (2-8)\times 10^{-7}$ \cite{rrs}, we note that the LD effects most likely dominate the exclusive decay rate.

In conclusion, we have calculated the long distance contributions to the exclusive $B\to K\gamma\gamma$ decay channel.  It was indicated that a mechnism which explains the observed large branching ratio for $\eta'$ production in charmless hadronic B decays leads to a significant resonance effect in this decay mode.  The combined intermediate pseudoscalar $\eta'$ and $\eta_c$ contributions are likely to dominate the decay rate, and therfore, to probe the short distance physics, one should look for regions in various decay distributions where these long distance effects are minimal.

\end{document}